\documentstyle[12pt,epsfig]{article}

\newlength{\dinwidth}
\newlength{\dinmargin}
\begin{document}
\def\bold#1{\setbox0=\hbox{$#1$}%
     \kern-.025em\copy0\kern-\wd0
     \kern.05em\%\baselineskip=18ptemptcopy0\kern-\wd0
     \kern-.025em\raise.0433em\box0 }
\def\slash#1{\setbox0=\hbox{$#1$}#1\hskip-\wd0\dimen0=5pt\advance
         to\wd0{\hss\sl/\/\hss}}
\newcommand{\be}{\begin{equation}}
\newcommand{\ee}{\end{equation}}
\newcommand{\bea}{\begin{eqnarray}}
\newcommand{\eea}{\end{eqnarray}}
\newcommand{\nn}{\nonumber}
\newcommand{\dd}{\displaystyle}
\newcommand{\bra}[1]{\left\langle #1 \right|}
\newcommand{\ket}[1]{\left| #1 \right\rangle}
\newcommand{\spur}[1]{\not\! #1 \,}
\thispagestyle{empty} \vspace*{1cm} \rightline{BARI-TH/04-484}
\vspace*{2cm}
\begin{center}
  \begin{LARGE}
  \begin{bf}
The Riddle of  Polarization \\ \vspace*{0.5cm}
in $B \to VV$ Transitions
 \vspace*{0.5cm}
  \end{bf}
  \end{LARGE}
\end{center}
\vspace*{8mm}
\begin{center}
\begin{large}
P. Colangelo$^a$,  F. De Fazio$^a$ and T.N.Pham$^b$
\end{large}
\end{center}
\begin{center}
\begin{it}
$^a\;\;$Istituto Nazionale di Fisica Nucleare, Sezione di Bari, Italy\\
$^b\;\;$Centre de Physique Th\'eorique, \\
Centre National de la Recherche Scientifique, UMR 7644\\
\'Ecole Polytechnique, 91128 Palaiseau Cedex, France\\
\end{it}

\end{center}
\begin{quotation}
\vspace*{1.5cm}
\begin{center}
  \begin{bf}
  Abstract\\
  \end{bf}
  \end{center}
\noindent
Measurements of polarization fractions in $B \to VV$ transitions,
with $V$ a light vector meson, show that the longitudinal amplitude dominates
in $B^0 \to \rho^+ \rho^-$,  $B^+ \to \rho^+ \rho^0$,
and $B^+ \to \rho^0 K^{*+}$ decays
and not in the penguin induced decays  $B^0 \to \phi K^{*0}$,
$B^+ \to \phi K^{*+}$. We study the effect of rescattering mediated by 
charmed resonances, finding that 
in $B \to \phi K^*$  it can be responsible of the suppression of the longitudinal amplitude.
For the decay $B \to \rho K^*$ we find that the longitudinal fraction cannot be too large
without invoking  new effects. 

\end{quotation}
\newpage
\baselineskip=18pt
\vspace{2cm}
\section{Introduction}\label{sec:intro}
An important result obtained by  Belle and BaBar Collaborations is
the measurement of the decay widths and of the polarization
fractions of several B decays to two light vector mesons
\cite{belle,Aubert:2003mm,Zhang:2003up,Aubert:2003xc}. The
branching fractions measured by the two Collaborations are
collected in Table~\ref{tab:tab1} together with the averages.
Together with these data one should  collect the upper bound
${\cal B}(B^0 \to \rho^0 \rho^0)\le 2.1 \times 10^{-6}$ from BaBar
\cite{Aubert:2003mm}. Through the analysis of  angular
distributions, the polarization fractions of the  final states
have been measured  as reported in
Table~\ref{tab:tab2}. In  the decay modes $B^0 \to \rho^+ \rho^-$
and  $B^+ \to \rho^0 \rho^+, \rho^0 K^{*+}$ the final states are
essentially in longitudinal configuration, with a larger uncertainty for $B^+ \to \rho^0 K^{*+}$; on the contrary, in
both the observed $B \to \phi K^*$ transitions the longitudinal
amplitude does not dominate, providing nearly $~50\%$ of the rate.
\begin{table}[ht]
\begin{center}
\vspace*{0.5cm}
\begin{tabular}{c c c c }\hline \hline
Mode&Belle~\cite{belle}&BaBar~\cite{Aubert:2003mm}&Average\\ \hline
$B^+\to\phi K^{*+}$&$(6.7^{+2.1+0.7}_{-1.9-1.0})\times10^{-6}$&$(12.7^{+2.2}_{-2.0}\pm1.1)\times10^{-6}$&$(9.5\pm1.7)\times10^{-6}$\\
$B^0\to\phi
K^{*0}$&$(10.0^{+1.6+0.7}_{-1.5-0.8})\times10^{-6}$&$(11.2\pm1.3\pm0.8)\times10^{-6}$&$(10.7\pm1.2)\times10^{-6}$\\
\hline \hline
Mode&Belle~\cite{Zhang:2003up}&BaBar~\cite{Aubert:2003mm,Aubert:2003xc}&Average\\ \hline
$B^+\to\rho^0 K^{*+}$&&$(10.6^{+3.0}_{-2.6}\pm2.4)\times10^{-6}$&\\
$B^+\to\rho^0 \rho^+$&$(31.7\pm7.1^{+3.8}_{-6.7})\times10^{-6}$&$(22.5^{+5.7}_{-5.4}\pm5.8)\times10^{-6}$&$(26.2\pm6.2)\times10^{-6}$\\
$B^0\to\rho^+ \rho^-$&&$(25^{+7+5}_{-6-6})\times10^{-6}$&\\
\hline\hline
\end{tabular}
\end{center}
\caption{Branching fractions of $B\to VV$ decay modes.}\label{tab:tab1}
\end{table}

There are  reasons to expect that the light VV final state should be
mainly longitudinally polarized, see, e.g., the discussion in 
 \cite{Kagan:2004uw}. 
In the following we summarize the arguments, which essentially rely
on factorization  and on the infinite heavy quark mass limit.
Invoking such arguments, the deviation observed in  $B\to \phi K^*$
could be interpreted as a signal of new physics  \cite{Grossman:2003qi}.
A more orthodox interpretation \cite{Kagan:2004uw},
 in the framework
of QCD improved factorization \cite{neubert}, relies on the observation
that (logarithmically divergent) annihilation diagrams can modify
the polarization amplitudes in $B \to \phi K^*$, producing fractions
in agreement with observation.

In this note we wish to address another effect that potentially changes
the result in the penguin induced $B \to \phi K^*$ decay
without affecting the observed $B \to \rho \rho$
transition: rescattering  of intermediate
charm states. Such effects, studied long ago in $B \to K \pi$ transitions
\cite{Colangelo:1989gi} and investigated  recently in other
$B \to PP$ and $VP$ decays
\cite{isola} as well as in factorization forbidden B transitions to charmonium
final states \cite{Colangelo:2002mj},
can invalidate the arguments on
the basis of which the dominance of the longitudinal configuration
is argued.

We discuss factorization and its consequences in Section~\ref{sec:fact}
and the analysis of rescattering effects  for $B^0 \to \phi K^{*0}$
in Section~\ref{sec:resc}.
At the end we discuss a few consequences.
\begin{table}[ht]
\begin{center}
\vspace*{0.5cm}
\begin{tabular}{c c c c c}\hline \hline
Mode&Pol. fraction&Belle~\cite{belle}&BaBar~\cite{Aubert:2003mm}& Average\\
\hline
$B^+\to\phi K^{*+}$&$\Gamma_L/\Gamma$&&$0.46\pm0.12\pm0.03$&\\
&&&&\\
$B^0\to\phi K^{*0}$&$\Gamma_L/\Gamma$&$0.43\pm0.09\pm0.04$&$0.65\pm0.07\pm0.02$&$0.58\pm0.06$\\
&&&$(0.52\pm0.07\pm0.02)$&\\
$B^0\to\phi K^{*0}$&$\Gamma_\perp/\Gamma$&$0.41\pm0.10\pm0.02$&$(0.27\pm0.07\pm0.02)$&\\
\hline \hline
Mode&Pol. fraction&Belle~\cite{Zhang:2003up}&BaBar~\cite{Aubert:2003mm,Aubert:2003xc}& Average\\
\hline
$B^+\to\rho^0 K^{*+}$&$\Gamma_L/\Gamma$&&$0.96^{+0.04}_{-0.15}\pm0.04$&\\
$B^+\to\rho^0 \rho^+$&$\Gamma_L/\Gamma$&$0.95\pm0.11\pm0.02$&$0.97^{+0.03}_{-0.07}\pm0.04$&$0.96\pm0.07$\\
$B^0\to\rho^+ \rho^-$&$\Gamma_L/\Gamma$&&$0.98^{+0.02}_{-0.08}\pm0.03$&\\
\hline \hline
\end{tabular}
\end{center}
\caption{Polarization fractions in $B\to VV$ transitions.
The BaBar results reported in brakets are preliminary data quoted
in ref.~\cite{gritsan}.}\label{tab:tab2}
\end{table}

\section{Polarization in factorization-based approaches}\label{sec:fact}
The decay $B^0 \to \phi K^{*0}$ is described by the amplitude \be
{\cal A}(B^0(p) \to \phi(q,\epsilon) K^{*0}(p^\prime, \eta))=
{\cal A}_0 \, \epsilon^*\cdot \eta^* + {\cal A}_2 \,
(\epsilon^*\cdot p) (\eta^* \cdot q)+ i {\cal A}_1 \,
\epsilon^{\alpha \beta \gamma \delta} \epsilon^*_\alpha
\eta^*_\beta p_\gamma p^\prime_\delta  \label{eq:amp}
\ee with
$\epsilon(q,\lambda)$ and $\eta(p^\prime,\lambda)$ the $\phi$ and
$K^*$ polarization vectors, respectively, with $\lambda=0,\pm1$
the three helicities.
Since the decaying $B$ meson is spinless, the final vector mesons
share the same helicity. ${\cal A}_0$ and ${\cal A}_2$ are
associated to the S- and D-wave decay, respectively,  and ${\cal
A}_1$ to the P-wave transition.

The three helicity amplitudes ${\cal A}_L$ and ${\cal A}_\pm$
can be written in terms of ${\cal A}_{0,1,2}$:
\bea
{\cal A}_L&=&-{1 \over M_\phi M_{K^*}}
[ (p \cdot p^\prime-M^2_{K^*}) {\cal A}_0+M_B^2 |\vec p^\prime|^2  {\cal A}_2]
\nn \\
{\cal A}_\pm&=&- {\cal A}_0\mp  M_B |\vec p^\prime|  {\cal
A}_1\,\,\,\,\ ; \eea in the transversity basis, the transverse
amplitudes \bea
{\cal A}_\parallel&=&{{\cal A}_+ +{\cal A}_- \over \sqrt 2}=-\sqrt 2 {\cal A}_0 \\
{\cal A}_\perp&=&{{\cal A}_+ - {\cal A}_- \over \sqrt 2}= - \sqrt
2 M_B |\vec p^\prime| {\cal A}_1 \,\,\, \nn \eea can also be
defined,  with $|\vec p^\prime|= \lambda^{1\over2}(M_B^2,
M_{K^*}^2, M_\phi^2) / 2 M_B$ ($\lambda$ the triangular function)
the common $\phi$ and $K^*$ three-momentum in the rest frame of
the decaying B-meson. In terms of such amplitudes the expression
of the decay rate is simply: \be \Gamma={|\vec p^\prime| \over 8 \pi
M_B^2} \big(|{\cal A}_L|^2+|{\cal A}_\parallel|^2+|{\cal
A}_\perp|^2\big) \,\,\,\, , \ee while  the three polarization fractions are
given by \bea f_L&=&{\Gamma_L \over \Gamma}= {\displaystyle |{\cal
A}_L|^2 \over
|{\cal A}_L|^2+|{\cal A}_\parallel|^2+|{\cal A}_\perp|^2 } \nn \\
f_\parallel&=&{\Gamma_\parallel \over \Gamma}= {\displaystyle
|{\cal A}_\parallel|^2 \over
|{\cal A}_L|^2+|{\cal A}_\parallel|^2+|{\cal A}_\perp|^2 } \\
f_\perp&=&{\Gamma_\perp \over \Gamma}= {\displaystyle |{\cal A}_\perp|^2 \over
|{\cal A}_L|^2+|{\cal A}_\parallel|^2+|{\cal A}_\perp|^2 }  \,\,\,\, . \nn
\eea

In order to compute the amplitude eq.(\ref{eq:amp}), we consider
the effective weak Hamiltonian inducing the $\bar b \to \bar s s
\bar s$ transitions, which can be written as \be H_W={G_F \over \sqrt 2}
(-V^*_{tb} V_{ts}) (\sum_{i=3}^{10} c_i {\cal O}_i + c_{7\gamma}
{\cal O}_{7\gamma} + c_{8g} {\cal O}_{8g}) \label{eq:ham} \ee with
the operators \bea
{\cal O}_3&=&(\bar b_\alpha s_\alpha)_{V-A} \sum_{q^\prime} (\bar q^\prime_\beta q^\prime_\beta)_{V-A} \nn \\
{\cal O}_4&=&(\bar b_\beta s_\alpha)_{V-A} \sum_{q^\prime} (\bar q^\prime_\alpha q^\prime_\beta)_{V-A} \nn \\
{\cal O}_5&=&(\bar b_\alpha s_\alpha)_{V-A} \sum_{q^\prime} (\bar q^\prime_\beta q^\prime_\beta)_{V+A} \nn \\
{\cal O}_6&=&(\bar b_\beta s_\alpha)_{V-A} \sum_{q^\prime} (\bar q^\prime_\alpha q^\prime_\beta)_{V+A} \\
{\cal O}_7&=&{3\over2}(\bar b_\alpha s_\alpha)_{V-A} \sum_{q^\prime} e_{q^\prime}(\bar q^\prime_\beta q^\prime_\beta)_{V+A} \nn \\
{\cal O}_8&=&{3\over2}(\bar b_\beta s_\alpha)_{V-A} \sum_{q^\prime} e_{q^\prime}(\bar q^\prime_\alpha q^\prime_\beta)_{V+A} \nn \\
{\cal O}_9&=&{3\over2}(\bar b_\alpha s_\alpha)_{V-A} \sum_{q^\prime} e_{q^\prime}(\bar q^\prime_\beta q^\prime_\beta)_{V-A} \nn \\
{\cal O}_{10}&=&{3\over2}(\bar b_\beta s_\alpha)_{V-A} \sum_{q^\prime} e_{q^\prime}(\bar q^\prime_\alpha q^\prime_\beta)_{V-A} \nn
\eea
($\alpha,\beta$  are colour indices and
$(\bar q q)_{V \mp A}=\bar q \gamma^\mu  (1 \mp \gamma_5) q$).
${\cal O}_{3-6}$ are gluon penguin operators,
${\cal O}_{7-10}$ electroweak penguin operators,
${\cal O}_{7\gamma}={ e \over 8 \pi^2} m_b \bar b \sigma^{\mu \nu} (1+\gamma_5) s
F_{\mu \nu}$ and
${\cal O}_{8g}={ g \over 8 \pi^2} m_b \bar b \sigma^{\mu \nu} (1+\gamma_5) T^a s
G^a_{\mu \nu}$, with  $F_{\mu \nu}$ and $G^a_{\mu \nu}$
the electromagnetic and the gluon field strength, respectively;
$c_{i,7\gamma,8g}(\mu)$ are the Wilson coefficients.

The amplitude ${\cal A}(B^0 \to \phi K^{*0})$ obtained from
(\ref{eq:ham}) admits a factorized form
\be
{\cal A}_{fact}(B^0 \to \phi K^{*0})= {G_F \over \sqrt 2}(-V^*_{tb} V_{ts}) a_W
\langle K^{*0}(p^\prime, \eta)|(\bar b s)_{V-A}|B^0(p)\rangle
\langle \phi(q,\epsilon)|(\bar s s)_V|0 \rangle
\label{eq:ampfact}
\ee
with $\displaystyle a_W=a_3+a_4+a_5-{1\over 2}(a_7+a_9+a_{10})$,
$\displaystyle a_i=c_i+{c_{i+1}\over N_c}$ for $i=3,5,7,9$ and
$\displaystyle a_i=c_i+{c_{i-1}\over N_c}$ for $i=4,10$ ($N_c$ is the number of colours).
This formula presents the drawbacks of  naive factorization,
namely there is not a compensation of the scale dependence between Wilson
coefficients and operator matrix elements.
However, it allows us to immediately write down the polarization
fractions, once the $B \to K^*$ matrix element has been expressed in
terms of form factors \footnote{For the $B \to K^*$ and $B\to D^*$ matrix elements 
eqs.(\ref{eq:BK}) and (\ref{eq:BD})  we use the same phase convention.} ,
and the $\phi$ meson leptonic constant  has been introduced:
\be
\langle \phi(q, \epsilon)|\bar s \gamma^\mu s|0 \rangle = f_\phi M_\phi
\epsilon^{*\mu}
\ee
\bea
\langle K^*(p^\prime, \eta)|\bar b \gamma_\mu (1-\gamma_5) s|B(p) \rangle &=&
-i \epsilon_{\mu \nu \rho \sigma} \eta^{*\nu} p^\rho p^{\prime \sigma}
{2 V \over M_B + M_{K^*}}  - \big [ (M_B + M_{K^*}) A_1 \eta^*_\mu
\nn\label{eq:BK} \\
&-& {A_2  \over M_B + M_{K^*}} (\eta^* \cdot p) \, (p +
p^\prime)_\mu -2 M_{K^*} {(A_3-A_0) \over q^2} (\eta^*\cdot p)
q_\mu \big] \,\,,\nn \\ \label{eq:formf} \eea with the form
factors $V, A_1, A_2, A_3$ and $A_0$ functions of $q^2$. From
(\ref{eq:ampfact}-\ref{eq:formf}) it is easy to write down the
polarization amplitudes and check that, for large values of $M_B$,
\bea
{\cal A}_L &\propto& M_B^3 [( A_1(M_\phi^2)-A_2(M_\phi^2))+{M_{K^*}\over M_B}( A_1(M_\phi^2)+A_2(M_\phi^2))] \nn \\
{\cal A}_\parallel &\propto& M_B A_1(M_\phi^2)  \\
{\cal A}_\perp &\propto& M_B V(M_\phi^2) \,\,\,\, ,\nn
\eea
expressions which determine the behaviour of the three amplitudes
once the parametric dependence  on the heavy quark mass of the form factors
close to the maximum recoil point has been established. In the limit
$M_B \to \infty$ and for $q^2=0$ such a dependence has been investigated
\cite{Charles:1998dr}
with the result that the three form factors $V,A_1$ and $A_2$ should be equal:
$A_2/A_1=V/A_1=1$. One therefore expects:
\be
{\Gamma_L\over\Gamma}\simeq 1 + {\cal O}({1 \over M_B^2})  \,\,\,\,\,\,\, , \,\,\,\,\,\, {\Gamma_\parallel\over\Gamma_\perp}\simeq 1
\,\,\,\,  \ee
regardless, in this scheme, of the Wilson and CKM coefficients.
Assuming generalized factorization, with the substitution of the Wilson
coefficients $a_i$ with effective
parameters $a_i^{eff}$,  it is eventually possible to reconcile
the branching ratio with the experimental measurement, but not to modify
the polarization fractions, since the dependence on the $a_i$ cancels out in the ratios.
Therefore, in order to explain the small ratio $\Gamma_L /\Gamma$ within
the Standard Model
one has to look either at the finite mass corrections,
or at effects beyond factorization.

%
\begin{figure}[ht]
\begin{center}
\mbox{\epsfig{file=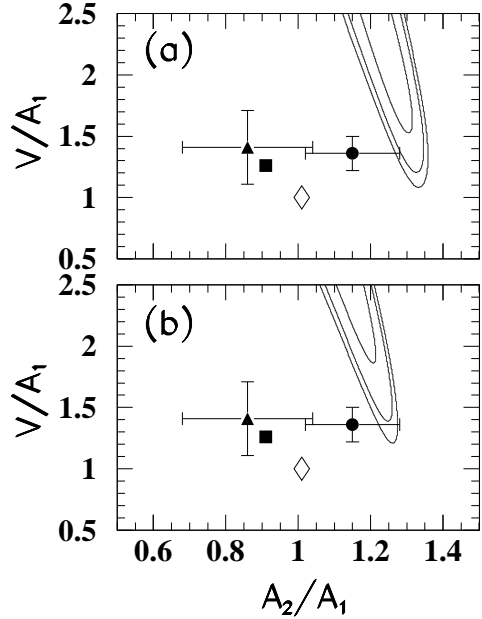, width=11cm}} \vspace*{-1.5cm}
\end{center}
\caption{ Ratios of $B \to K^*$ form factors:
$V(M^2_\phi)/A_1(M^2_\phi)$ versus  $A_2(M^2_\phi)/A_1(M^2_\phi)$.
The continuous lines correspond to the (one, two and
three-$\sigma$) regions of the Belle data in Table~\ref{tab:tab2}
(a) and of the average of Belle and BaBar data (b) for
$\Gamma_L/\Gamma$ and $\Gamma_\perp / \Gamma$ in $B^0 \to \phi
K^{*0}$. The points correspond to different form factor models:
QCDSR \cite{Colangelo:1995jv} (dot), LCSR \cite{Ball:2003rd}
(triangle), MS \cite{Melikhov:2000yu} (square), BSW
\cite{Bauer:1986bm} (diamond).} \vspace*{1.0cm} \label{fig:fit}
\end{figure}

For finite heavy quark mass,  one can compare the experimental result for the
polarization fractions in $B^0 \to \phi K^{*0}$ decays (Table~\ref{tab:tab2})
with the predictions of various form factor models
\cite{Colangelo:1995jv,Ball:2003rd,Melikhov:2000yu,Bauer:1986bm}. As shown
in fig.~\ref{fig:fit},  in many  models the ratios $A_2/A_1$ and $V/A_1$
deviate from the
asymptotic prediction, suggesting that  the regime of finite $M_B$
does not  fully coincide with the asymptotic regime.
In one case there is a marginal agreement
between the form factor model and data. However, the indication of effects
beyond  naive and generalized factorization is clear.

\section{Rescattering effects}\label{sec:resc}
\begin{figure}[ht]
\begin{center}
\vspace*{-1cm}
\hspace*{2cm}\mbox{\epsfig{file=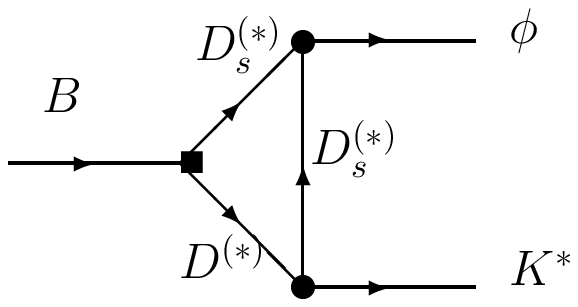, width=8cm}}
\vspace*{-2.5cm}
\end{center}
\caption{Rescattering diagrams contributing to
$B \to \phi K^*$. The box represents a weak
vertex, the dots strong couplings.} \vspace*{1.0cm}
\label{diagrams}
\end{figure}

If one considers  the possibility of rescattering effects, 
there are other terms in the effective weak hamiltonian that can  induce the  transition
$B^0 \to \phi K^{*0}$. Processes that should be  the most relevant ones are
 $\bar b \to c \bar c \bar s \to s \bar s \bar s$.
Such processes can give sizeable contribution to the penguin amplitudes obtained from
(\ref{eq:ham}) since they involve  Wilson coefficients of  ${\cal O}(1)$
(that multiply current-current quark operators), while the Wilson
coefficients in penguin
$\bar b \to \bar s s \bar s $ operators are smaller $({\cal O}(10^{-2}))$. On the other hand,
there is not a CKM suppression
in such processes, since  $|V^*_{tb} V_{ts}|$ and $|V^*_{cb} V_{cs}|$
are nearly equal.
An example of  processes of this type is depicted in fig.~\ref{diagrams},
where a sample of intermediate charm mesons  is  shown.
%
%
%
As far as the polarization of the final state is concerned, one
has to notice that different intermediate states in
fig.~\ref{diagrams} contribute to different polarization
amplitudes, so that the longitudinal as well as the transverse amplitudes  can be modified.
For example, considering only intermediate pseudoscalar and vector
charmed mesons coming from the $B$ meson vertex, there are eight  diagrams 
of the kind depicted  in fig.(\ref{diagrams}).
Intermediate states comprising one vector and one pseudoscalar
meson (four diagrams) only contribute to the $P-$wave transition
and therefore to the amplitude $\cal A_\perp$. On the
other hand, intermediate states comprising two pseudoscalar mesons
(two diagrams) only contribute to ${\cal A}_L$ and ${\cal
A}_\parallel$, while intermediate states with two vector mesons (2
diagrams) contribute to the three polarization amplitudes ${\cal
A}_L$,  $\cal A_\perp$
 and ${\cal A}_\parallel$.

In order to estimate the contribution of diagrams of the type  in
fig.\ref{diagrams} we can use a formalism that accounts for the
heavy quark spin-flavour symmetries in hadrons containing a single
heavy quark \cite{review}
and for the  so called hidden gauge symmetry to
describe their interaction with light vector mesons \cite{Casalbuoni:1992gi}.
As well
known, in the heavy quark limit, due to the decoupling of the
heavy quark spin $\vec s_Q$ from the light degrees of freedom
total angular momentum $\vec s_\ell$,  it is possible to classify
hadrons with a single heavy quark $Q$ in terms of  $s_\ell$.
Mesons can be collected in doublets the members of which only
differ for the relative orientation of $\vec s_Q$ and $\vec
s_\ell$ \cite{review}. The doublets with $J^P=(0^-,1^-)$
corresponding to $s_\ell^P= {1\over 2}^-$ can be described by the
effective fields 
\begin{equation}
H_a =
\frac{(1+{\rlap{v}/})}{2}[P_{a\mu}^*\gamma^\mu-P_a\gamma_5]
\label{neg}
\end{equation}
where $v$ is the meson four-velocity  and $a$ is a light quark
flavour index. 
The field $\overline H_a$ is  defined as
$\overline H_a= \gamma^0 H^\dagger_a \gamma^0$;
all the heavy field operators contain a factor $\sqrt{M_H}$ and
have dimension $3/2$.

It is possible to formulate an effective
Lagrange density for  the low energy interactions of heavy
mesons with light vector mesons \cite{Casalbuoni:1992gi}.
The interaction term of such a Lagrangian reads as
\begin{equation}
{\cal L}_{HHV} =
- i\; \beta \; Tr\{ H_b (v^\mu \rho_\mu)_{ba}  {\overline H}_a \}
\; + \; i\; \lambda \;Tr\{ H_b (\sigma^{\mu\nu} F_{\mu\nu})_{ba}
{\overline H}_a \} \,\,\ .
\label{L}
\end{equation}
Light vector mesons are included in (\ref{L}) through the fields
$\displaystyle \rho= i {g_V\over 2} \hat\rho$ representing the
low-lying vector octet:
\begin{equation}
{\hat \rho}=
\left (\begin{array}{ccc}
\sqrt{\frac{1}{2}}\rho^0+\sqrt{\frac{1}{6}}\omega_8 & \rho^+ &K^{*+}\nonumber\\
\rho^-&-\sqrt{\frac{1}{2}}\rho^0+\sqrt{\frac{1}{6}}\omega_8 &K^{*0}\\
K^{*-} & {\overline K}^{*0} &-\sqrt{\frac{2}{3}}\omega_8
\end{array}\right ) \label{M}
\end{equation}
with
$F_{\mu\nu}=\partial_\mu \rho_\nu-\partial_\nu \rho_\mu+[\rho_\mu, \rho_\nu]$.
Invoking the mixing $\omega_8-\omega_0$
one gets the interaction term involving $\phi$.
The  interactions of  heavy mesons
with the light vector mesons are
thus governed, in the heavy quark limit,  by two
couplings $\beta$ and $\lambda$.
From light cone QCD sum rules  \cite{Aliev:xb} as well as
 from vector mesons dominance arguments \cite{isola}
 one estimates
$\beta\simeq0.9$ and $\lambda\simeq 0.56$ GeV$^{-1}$, while $g_V$
is fixed to $g_V=5.6$  by the KSRF relation \cite{KSRF}.

Using (\ref{L}) it is easy to work out the matrix
elements $D^{(*)}_s D^{(*)} K^*$ appearing in one of the vertices
in fig.\ref{diagrams}:
\bea
\langle D_s^-(p_D-p^\prime) K^{*0}(p^\prime, \eta)|D^-(p_D=M_D v_D) \rangle &=&
\tilde \beta  \sqrt{M_D M_{D_s}} \; (v_D \cdot \eta^*) \nn \\
\langle D_s^{*-}(p_D-p^\prime,\epsilon_1) K^{*0}(p^\prime, \eta)|D^-(p_D) \rangle &=&
i \tilde \lambda \sqrt{M_D M_{D^*_s}} \;
\epsilon^{\alpha \nu \mu \beta} \; v_{D\alpha}  \eta^*_\nu p^\prime_\mu \epsilon^*_{1\beta} \nn \\
\langle D_s^-(p_D-p^\prime) K^{*0}(p^\prime, \eta)|D^{*-}(p_D,\eta_1) \rangle &=&
i \tilde \lambda\sqrt{M_{D^*} M_{D_s}} \;
\epsilon^{\alpha \nu \mu \beta} \; v_{D\alpha}  \eta^*_\nu p^\prime_\mu \eta_{1\beta} \nn \\
\langle D_s^{*-}(p_D-p^\prime,\epsilon_1) K^{*0}(p^\prime, \eta)|D^{*-}(p_D,\eta_1) \rangle &=&   - \tilde \beta
\sqrt{M_{D^*} M_{D^*_s}}\; (v_D\cdot\eta^*)\;(\epsilon_1^*\cdot\eta_1)\nn\\
+{\displaystyle  \tilde \lambda} \sqrt{M_{D^*} M_{D^*_s}} \;
[  (\eta_1\cdot\eta^*)\;(\epsilon_1^*\cdot p^\prime)&-&
(\eta_1\cdot p^\prime)\;(\epsilon_1^*\cdot \eta^*)] \nn \\
\label{eq:matrixel}
\eea
where
$\tilde \beta={\displaystyle 2  \beta g_V \over \sqrt2}$ and
$\tilde \lambda={\displaystyle 4  \lambda g_V \over \sqrt 2}$.
Matrix elements involving $\phi$
in the other vertex in fig.\ref{diagrams} are obtained analogously.

As for the weak amplitude  $B^0 \to D_s^{(*)+} D^{(*)-}$, since
there is empirical evidence
that factorization reproduces the main experimental
findings \cite{Luo:2001mc}, we write it as
\begin{equation}
\langle D_s^{(*)+} D^{(*)-} | H_W | B^- \rangle =
\displaystyle{G_F \over \sqrt{2}}V_{cb}V_{cs}^* a_1
\langle D^{(*)-} | (V-A)^\mu | B^0 \rangle
\langle D_s^{(*)+}| (V-A)_\mu | 0 \rangle
\label{fact}
\end{equation}
with $a_1\simeq 1$.
In the heavy quark limit  the  matrix elements in (\ref{fact})
involve the Isgur-Wise function \cite{review}:
\begin{eqnarray}
\langle D^-(v^\prime)|V^\mu|B^0(v)\rangle&=&\sqrt{M_B M_D} \; \xi(v \cdot v^\prime)
(v+v^\prime)^\mu\nonumber \\
\langle D^{*-}(v^\prime,\epsilon)|V^\mu|B^0(v)\rangle&=& - i
\sqrt{M_B M_{D^*}} \; \xi(v \cdot v^\prime) \; \epsilon^*_{\beta} \;
\varepsilon^{\alpha \beta \gamma \mu} v_\alpha v^\prime_\gamma \label{BD} \label{eq:BD}\\
\langle D^{*-}(v^\prime,\epsilon)|A^\mu|B^0(v)\rangle&=&
\sqrt{M_B M_{D^*}} \; \xi(v \cdot v^\prime) \; \epsilon^*_{\beta}
[(1+v \cdot v^\prime) g^{\beta \mu}- v^\beta v^{\prime\mu}] \,\,\, ,\nonumber 
\end{eqnarray}
$v$ and $v^\prime$ being $B$ and $D^{(*)}$ four-velocities,
$\epsilon$ the $D^*$ polarization vector and
$\xi(v \cdot v^\prime)$ the Isgur-Wise form factor.
As for the $D^{(*)}$ current-vacuum matrix elements defined as
\begin{eqnarray}
\langle0|\bar q_a \gamma^\mu \gamma_5 c|D_a(v)\rangle
&=&f_{D_a} M_{D_a} v^\mu \nonumber
\\
\langle0|\bar q_a \gamma^\mu c|D^*_a(v,\epsilon)\rangle
&=&f_{D^*_a} M_{D^*_a} \epsilon^\mu \,\,\, ,
\end{eqnarray}
they can be  parameterized in the heavy quark limit in terms of a single
quantity $f_{D_a}=f_{D^*_a}$.

Now, the estimate of the absorptive part of the rescattering diagrams in fig.~\ref{diagrams}
\be
{\rm Im}{\cal A}_{resc}=
{{\lambda^{1\over2}(M_B^2,M_{D_s^{(*)}}^2,M_{D^{(*)}}^2)}\over32 \pi M_B^2}
\int_{-1}^{+1} dz {\cal A}(B^0 \to D_s^{(*)+} D^{(*)-})
 {\cal A}(D_s^{(*)+} D^{(*)-}\to \phi K^{*0}) \label{amp}
\ee
can be carried out. The integration variable $z=\cos \theta$ is related to the angle between
the three-momenta of $\phi$ and of the emitted  $D^{(*)}_s$ from $B$ vertex in fig.\ref{diagrams}.
We use
$|V_{cb}|=0.042$, $|V_{cs}|=0.974$ (the central
values reported by the Particle Data Group \cite{Hagiwara:fs}),
$f_{D^*_s}=f_{D_s}=240$ MeV \cite{Colangelo:2000dp}
and  $\xi(y)= \left( {2 \over 1+y} \right)^2$.

The couplings in (\ref{eq:matrixel})
do not account for the off-shellness of the exchanged
$D^{(*)}_{s}$  mesons in fig.\ref{diagrams}.
One can introduce  form factors:
\begin{equation}
g_i(t)=g_{i0}\,F(t)\,, \label{offshell}
\end{equation}
to account for the $t$-dependence of the couplings
(the vertices in rescattering diagrams cannot be considered
point-like since they do not involve elementary particles),
$g_{i0}$ being the on-shell couplings. However,
the form factors are unknown. We use
\be
F(t)=\displaystyle{\Lambda^2 -M^2_{D^*_s} \over \Lambda^2-t}
\label{ff}\ee
to satisfy QCD counting rules.
We could vary the value of $\Lambda$, considering the
uncertainty from the form factor $F(t)$
in the final numerical result. Instead, since
the relative sign of rescattering and factorized amplitude is also unknown,
as well as the role of diagrams
involving excitations and the continuum, we fix $\Lambda=2.3$ GeV and analyze
the sum
\be
{\cal A}={\cal A}_{fact} + r {\cal A}_{resc}
\label{eq:sumamp}
\ee
varying the parameter $r$ and approximating the long distance amplitude with eq.(\ref{amp}).

\begin{figure}[ht]
\begin{center}
\mbox{\epsfig{file=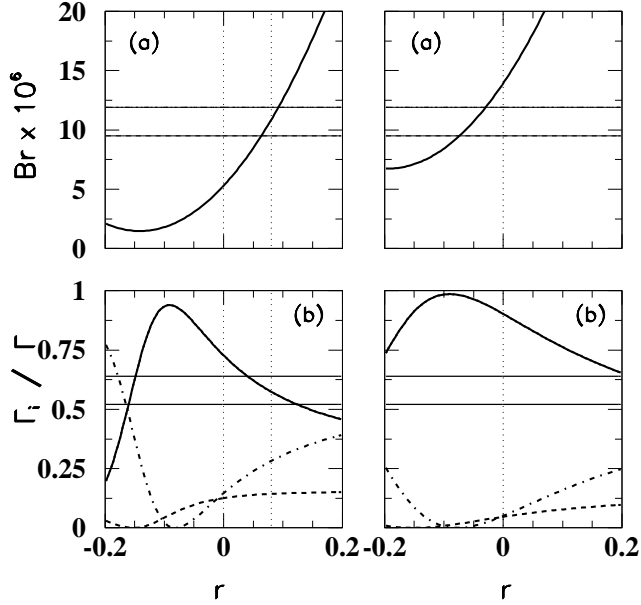, width=10cm}}
\vspace*{-1.0cm}
\end{center}
\caption{Dependence of the branching ratio
and polarization fractions
of $B^0 \to \phi K^{*0}$ on the long distance contribution.
$B\to K^*$ form factors computed in \cite{Colangelo:1995jv} (left) and
in \cite{Ball:2003rd} (right) are used in the short-distance amplitude.
$r=0$ corresponds to the absence of rescattering.
 The three curves in (b)
correspond to
$\Gamma_L/\Gamma$ (continuous curve),
$\Gamma_\perp/\Gamma$ (dashed) and
$\Gamma_\parallel/\Gamma$ (dot-dashed). The horizontal lines  represent the experimental
result in Table~\ref{tab:tab1} for the branching ratio (a) and for
$\Gamma_L/\Gamma$ (b).}
\vspace*{1.0cm}
\label{fig:rate}
\end{figure}
We compute the short-distance factorized amplitude
using the $B \to K^*$ form factors appearing in two extreme cases in
fig.\ref{fig:fit}, the model  \cite{Colangelo:1995jv} and
the model \cite{Ball:2003rd}, with
 Wilson coefficients 
$a_3=48 \times 10^{-4}$,
$a_4=(-439 -77 i) \times 10^{-4}$,
$a_5=-45 \times 10^{-4}$,
$a_7=(-0.5-1.3 i) \times 10^{-4}$,
$a_9=(-94-1.3 i)48 \times 10^{-4}$ and
$a_{10}=(-14-0.4 i) \times 10^{-4}$,  as computed  in \cite{Ali:1998eb} for $N_c=3$.

The result is depicted in fig.\ref{fig:rate}.
For the model  \cite{Colangelo:1995jv},
a contribution of the rescattering amplitude is in order to obtain the
measured $B \to \phi K^*$ branching fraction. Of the two possible values of
the parameter $r$ which reproduce the experimental rate, $r\simeq 0.08$
and $r\simeq-0.3$,  the former
allows us to simultaneously obtain a small longitudinal polarization
fraction:
$\displaystyle {\Gamma_L / \Gamma} \simeq 0.55$,
compatible with the measurements.
The tranverse polarization fractions turn out
$\displaystyle{{\Gamma_\parallel / \Gamma} \simeq 0.30}$  and
$\displaystyle{\Gamma_\perp / \Gamma} \simeq 0.15$. They are both consistent
with measurement, but with the hierarchy
$\displaystyle {\Gamma_\parallel / \Gamma} > {\Gamma_\perp / \Gamma}$.

If we use the form factors in
\cite{Ball:2003rd},  for $r=0$ the predicted rate exceeds
the experimental datum, so that the rescattering contribution should
be weighted by a negative  $r$   to reconcile the branching fraction;  
as depicted in fig.\ref{fig:rate}, in such a region ($r\simeq-0.05$)
the longitudinal fraction increases.  However, this conclusion crucially  depends on the value of the
Wilson coefficients $a_3 - a_{10}$ used as an input in the evaluation
of the short-distance amplitude.
As shown in  \cite{Ali:1998eb} , for example, $a_4$ varies from $-402-72i$ to $-511-87i$
changing $N_c$ from $2$ to $\infty$. 
For a smaller value of the sum of Wilson coefficients, both the sets of form factors
would require a similar long-distance contribution, with the effect of reducing
the longitudinal fraction. 

A  feature of both the sets of data is that, in the
region of $r$ where the experimental rate is reproduced,
$\Gamma_\parallel$ is larger or similar to  $\Gamma_\perp$. The
ratio $\displaystyle{\Gamma_\parallel \over \Gamma_\perp}$ is
sensitive to operators of different chirality which would  appear
in the effective Hamiltonian in extensions of the Standard Model
\cite{Kagan:2004uw}.

\section{Discussion}

The conclusion of this  analysis is that FSI effects can  modify
the helicity amplitudes in penguin dominated  processes. 
The numerical result depends on the interplay between  Wilson coefficients,
form factors and rescattering amplitude, and we have shown that
the experimental observation  can be  reproduced.
At the
same time, the rescattering effects we have considered are too small to  affect the observed $B \to \rho
\rho$ decays. As a matter of fact, while the CKM factors in the
tree diagram in $B^0 \to \rho^+ \rho^-$
transition ($V^*_{ub}V_{ud}$) have similar size to the CKM
factor in the FSI diagram in fig.\ref{diagrams}
($V^*_{cb}V_{cd}$), the Wilson coefficient in  current-current
transition is ${\cal O}(1)$.
We can expect to observe FSI effects in colour-suppressed and  other
penguin induced $B\to VV$ decays, such as
${ B}^0 \to \rho^0  K^{*0}$,
${B}^0 \to \omega  K^{*0}$, and
${B}^0 \to \rho^0 \rho^{0}$,
${ B}^0 \to \rho^0 \omega$,
$B^- \to \rho^- \overline K^{*0}$,
$B^- \to K^{*-} K^{*0}$.

Let us consider  $B^+ \to \rho^0 K^{*+}$. On the basis of general
arguments, we cannot  assess the role of   FSI 
without an explicit calculation, due to the CKM suppression of the
factorized amplitude.
The determination of the rescattering amplitude,  similar to that in fig.\ref{diagrams},
can be done following the same method discussed above, obtaining 
 $\displaystyle \Gamma_L / \Gamma \simeq 0.7$, i.e. smaller (even though
compatible within 2-$\sigma$ ) than the measurement in Table \ref{tab:tab2}.

Therefore, in our approach  we can accomodate
a small $\Gamma_L$ for  $B \to \phi K^*$  at the prize of having a
smaller value
of  $\Gamma_L$ for  $B \to \rho K^*$,  which is not currently excluded
due to the  uncertainty in the data for this mode.  It is interesting to
notice that an analogous prediction is done in QCD improved factorization
\cite{Kagan:2004uw}, where one gets
$\displaystyle {\Gamma_L \over \Gamma}(B\to \rho K^*)
<{\Gamma_L \over \Gamma}(B\to \phi K^*)$.
More precise measurements are in order to suggest a solution to this polarization riddle.
If  further  measurements
of  polarization fractions will confirm the present situation of a
small longitudinal fraction in $\phi K^*$ and a large longitudinal fraction
in  $\rho K^*$, in that case  we cannot identify
uniquely the rescattering mechanism for explaining the data, envisaging the exciting
necessity of new effects.

\vspace*{1cm}
\noindent {\bf Acnowledgments.}
One of us (PC) would like to thank 
CPhT,  \'Ecole Polytechnique, where this work was completed.
We acknowledge partial support from the EC Contract No.
HPRN-CT-2002-00311 (EURIDICE).

\end{document}